\documentclass[pre,showpacs,twocolumn,amsmath,superscriptaddress,
preprintnumbers,aps]{revtex4-1}

\usepackage{amssymb}
\usepackage{graphicx}

\begin{document}

\title{Wave systems with direct processes and localized losses or gains: the
non-unitary Poisson kernel}

\author{A. M. Mart\'inez-Arg\"uello} 
\affiliation{Instituto de Ciencias F\'isicas, Universidad Nacional Aut\'onoma
de M\'exico,  Apartado Postal 48-3, 62210 Cuernavaca Mor., Mexico}

\author{R. A. M\'endez-S\'anchez}
\affiliation{Instituto de Ciencias F\'isicas, Universidad Nacional Aut\'onoma
de M\'exico,  Apartado Postal 48-3, 62210 Cuernavaca Mor., Mexico}

\author{M. Mart\'inez-Mares}
\altaffiliation{Permanent address: Departamento de F\'{\i}sica, Universidad
Aut\'onoma Metropolitana-Iztapalapa, Apartado Postal 55-534, 09340 M\'exico D.
F., Mexico. E-mail: moi@xanum.uam.mx}
\affiliation{Instituto de Ciencias F\'isicas, Universidad Nacional Aut\'onoma
de M\'exico, Apartado Postal 48-3, 62210 Cuernavaca Mor., Mexico}


\begin{abstract}
We study the scattering of waves in systems with losses or gains simulated by
imaginary potentials. This is done for a complex delta potential that
corresponds to a  spatially localized absorption or amplification. In the Argand
plane the scattering matrix moves on a circle $C$ centered on the real axis, but
not at the origin, that is tangent to the unit circle. From the numerical
simulations it is concluded that the distribution of the scattering matrix, when
measured from the center of the circle $C$, agrees with the non-unitary Poisson
kernel. This result is also obtained analytically by extending the analyticity
condition, of unitary scattering matrices, to the non-unitary ones. We use this
non-unitary Poisson kernel to obtain the distribution of non-unitary scattering
matrices when measured from the origin of the Argand plane. The obtained
marginal distributions have an excellent agreement with the numerical results.
\end{abstract}


 
\maketitle


\section{Introduction}

In recent years there has been an intense study of several properties of
classical wave chaotic systems, such as microwave 
cavities~\cite{Graf,Sridhar,Kuhl} and graphs~\cite{Hul}, elastic, acoustic, and
optical resonators~\cite{Arcos,Gutierrez,Schaadt,Doya}. These systems inevitably
exhibit energy dissipation (absorption) and many experimental and theoretical
investigations have focused on the effect of this absorption on the scattering
properties (see Refs.~\cite{Kuhl,JETPLett,Fyodorov,Baez} and references
therein).
 
In addition to losses of energy the scattering properties are also affected by
the imperfect coupling of the antennas to the system, that gives rise to a
prompt response due to direct reflections~\cite{Rafael1}. It has been shown that
for chaotic systems the distribution of the \emph{sub-unitary} scattering
matrix, $\tilde{S}$, is modulated by a generalization of the Poisson kernel,
Poisson's kernel squared, in a single port, or one channel,
configuration~\cite{Rafael2}. For an arbitrary number of channels, it was found
that the Poisson kernel squared is the Jacobian of the transformation between a
non-unitary scattering matrix with direct processes and a non-unitary one
without such processes~\cite{Gopar}. 

The Poisson kernel was developed first for \emph{unitary} scattering matrices.
In the one channel (one-dimensional) case, it is obtained from the analytical
structure of the scattering matrix $S$ and can be interpreted as the probability
to find $S$ in a certain region of its space~\cite{Lopez,LesHouches,MelloKumar}.
For an arbitrary number of channels the Poisson kernel is the probability
density of $S$ with maximum information entropy, where the only physically
relevant parameter is the energy average $\overline{S}$, known as the
\emph{optical $S$ matrix}. In the optical model~\cite{Feshbach} $\overline{S}$
quantifies the direct processes~\cite{MelloKumar}; in the absence of such
processes, $\overline{S}=0$ and the probability density of $S$ is a constant. 

In a stationary random processes the analyticity-ergodicity conditions yield the
Poisson kernel as the probability distribution of an ensemble of $S$ matrices
determined by the ensemble average $\langle S\rangle$~\cite{MelloKumar}. This
ensemble represents an ensemble of macroscopically identical systems 
that describes the statistical fluctuations of scattering properties in chaotic
cavities with direct processes~\cite{Baranger1,Baranger2}. When 
$\langle S\rangle=0$, $S$ is uniformly distributed. Also, the Poisson
kernel appears as the Jacobian of a transformation between scattering matrices,
with and without direct processes~\cite{Friedman}. A physical realization of
that transformation was given in Ref.~\cite{Brouwer}; this realization was helpful to demonstrate the equivalence, in certain limit, between the imaginary potential model and the voltage probe model, both used to implement dephasing or absorption in ballistic chaotic cavities~\cite{Brouwer1997}. 

Our purpose in this paper is to deep in the understanding of the Poisson kernel.
We study the non-unitary scattering matrix, $\tilde{S}$, in a one-dimensional
(1D) problem with spatially localized losses or gains, in the presence of promt
responses. We analyze the analytical structure of $\tilde{S}$ to find its
probability of stay in a certain region of its space. It is worth mentioning
that local losses or gains have been of theoretical and experimental interest
in chaotic systems. In this way, local losses have been used to model surface
absorption in chaotic cavities~\cite{Surface}; quasimodes in elastic cavities
were measured when the absorption increase in a progresive way~\cite{Xeridat}.
Also, local gains were used to select scar modes in multimode
cavities~\cite{Michel}. Furthermore, a waveguide terminated by a perfect
absorber, was used to study fidelity in chaotic microwave
cavities~\cite{Koeber}. By the other hand, the nearest level spacing statistics
were investigated in open chaotic systems, as a function of the coupling to the
environment~\cite{Poli}. However, we recall that in the present work we leave
chaos behind and we concentrate in the local character of the absorption or
amplification.

We organize the paper as follows. In the next section we use the analytical
structure in the unitary case to get the analyticity condition for non-unitary
scattering matrices, in the one channel situation. This is done by adding an
imaginary part to the energy. Also, we apply that analyticity condition to
non-unitary scattering matrices with constant modulus to obtain the Poisson
kernel. In Sec.~\ref{sec:1D} we consider a 1D problem with local absorption or
amplification and find the probability of stay of the scattering matrix
$\tilde{S}$ that describes it. We present our conclusions in
Sec.~\ref{sec:conclusions}

\section{Analyticity condition and Poisson's kernel}

The scattering matrix $S(E)$ of systems where the flux is conserved is a
unitary one that depends on the energy $E$. In the one channel case it is a
$1\times 1$ matrix that moves on the unit circle as $E$ is varied, but it does
not visit all points with the same frequency when the energy average
$\overline{S(E)}$ of $S(E)$, does not vanish. Using the fact that $S(E)$ is an
analytic function in the upper half of the complex-$E$ plane, it is shown that
the energy average of the $m$-th power of $S(E)$ coincides with the $m$-th power
of $\overline{S(E)}$~\cite{MelloKumar}, 
\begin{equation}
\label{eq:condition1}
\overline{S^m(E)} = {\overline{S(E)}\,}^m.
\end{equation}
This result is known as analyticity condition.

Dissipation or amplification can be modeled by adding an imaginary part to the
energy~\cite{Doron,Brouwer1997}, in which case the scattering matrix, denoted by
$\tilde{S}(E)$, is no longer unitary, it is sub-unitary or over-unitary,
respectively. That is, $\tilde{S}(E)$ can be obtained from an unitary $S(E)$ by
extending the energy to the complex plane, 
\begin{equation}
\label{eq:analyticity1}
\tilde{S}(E) = S(E\pm iV_0) ,
\end{equation}
where $V_0>0$ and the plus (minus) sign holds for absorption (amplification).
Using Eqs.~(\ref{eq:condition1}) and~(\ref{eq:analyticity1}) it is easy to prove
that 
\begin{equation}
\label{eq:analyticity2}
\overline{{\tilde{S}}^m(E)} = {\overline{\tilde{S}(E)}\,}^m.
\end{equation}
This means that, in the non-unitary case, the scattering matrix satisfies the
same analyticity condition as in the unitary one.

In the $1\times 1$ case $\tilde{S}$ is a complex number that can be written in
a polar form as 
\begin{equation}
\label{eq:polar}
\tilde{S} = \sqrt{R}\, e^{i\theta},
\end{equation}
where $R$ is the reflection coefficient; $R<1$ for absorption and $R>1$ for
amplification. $R$ and $\theta$ vary with $E$ and, therefore, $\tilde{S}$ moves
in the Argand plane in the unit disk in presence of absorption, or outside of it
for amplification. Following Ref.~\cite{MelloKumar}, we assume the existence of
the measure 
\begin{equation}
dP(\tilde{S}) = p(R,\theta) \, dR\, d\theta ,
\end{equation}
which is the probability to find $R$ between $R$ and $R+dR$ and $\theta$
between $\theta$ and $\theta+d\theta$. 

A very particular situation is the one in which the non-unitary scattering
matrix moves along the circle with constant radius, as a function of $E$; we
denote this $1\times 1$ matrix by $\tilde{S}'=\sqrt{R'}e^{i\theta'}$, where $R'$
takes the fixed value $R_0$. Assuming that the energy average of $\tilde{S}'$ is
not null, we want to determine the probability to find $\theta'$ in the interval
between $\theta'$ and $\theta'+d\theta'$. In this case, 
\begin{equation}
dP(\tilde{S}') = p'(R', \theta')\, dR'\, d\theta'  
\end{equation}
with 
\begin{equation}
\label{eq:kernel}
p'(R', \theta') = \delta(R'-R_0) \, q'(\theta').
\end{equation}
This can be used to evaluate the average of the $m$th power of $\tilde{S}'$ as
follows,
\begin{equation}
\overline{{\tilde{S'}}^m} = 
\int {\tilde{S'}}^m dP'(\tilde{S}') = R_0^{m/2} \int_{0}^{2\pi} e^{im\theta'}
q'(\theta') d\theta' .
\end{equation}
Since $q'(\theta')$ is a periodic function of $\theta'$ it can be expanded in a
Fourier series, 
\begin{equation}
\label{eq:series}
q'(\theta') = \sum_{m=-\infty}^{\infty} c_{m} e^{im\theta'},
\end{equation}
where the $c_m$'s are constants that satisfy $c_{m}=c^*_{-m}$ to ensure
$q'(\theta')$ to be real. Using the analyticity
condition~(\ref{eq:analyticity2}) it is easy to see that
\begin{equation}
c_{-m} = \frac{1}{2\pi} \frac{{\overline{\tilde{S}'}\,}^m}{R_0^{m/2}} .
\end{equation}
With these coefficients the sum in Eq.~(\ref{eq:series}) can be done to obtain
\begin{equation}
\label{eq:Poisson-q}
q'(\theta') = \frac{1}{2\pi} 
\frac{R_0-\left|\overline{\tilde{S}'} \right|^2}
{\left|\tilde{S}' - \overline{\tilde{S}'} \right|^2}, 
\end{equation}
and, therefore, 
\begin{equation}
\label{eq:Poisson-Rq}
p'(R',\theta') = \frac{1}{2\pi} 
\frac{R'-\left|\overline{\tilde{S}'} \right|^2}
{\left|\tilde{S}' - \overline{\tilde{S}'} \right|^2} \,
\delta\left(R'-R_0\right).
\end{equation}
Equation~(\ref{eq:Poisson-q}) is the same expression as that obtained for a
disk of radius $R_0$ in the problem of transfer of heat~\cite{Krantz}.
Eq.~(\ref{eq:Poisson-Rq}) reduces to the Poisson's kernel of
Ref~\cite{MelloKumar} in the unitary case when we integrate it over the variable
$R$ for $R_0=1$. 

\section{Local absorption (amplification) in 1D cavities}
\label{sec:1D}

A simple model of a cavity with absorption or amplification and direct
reflection consists of a Dirac delta potential with complex intensity, located
at a distance $a$ in front of an impenetrable barrier. The potential is 
\begin{equation}
\label{eq:potential}
V(x)=\left\{ 
\begin{array}{cl}
\infty, & x<0 \\ 
(u \mp iv)\,\delta(x-a), & x>0
\end{array} 
\right. ,
\end{equation}
where $u$ and $v$ are positive constants; the minus (plus) sign corresponds to
absorption (amplification). Notice the local character of the absorption or
amplification. The incident waves to the potential suffer multiple scattering
before they leave the cavity formed between the infinite barrier and the delta
potential. The outgoing plane wave amplitude is related to the incoming one by
the $1\times 1$ scattering matrix given by
\begin{equation}
\tilde{S} = -\frac{[\sin ka + (k/\tilde{\alpha})\cos ka] + 
i(k/\tilde{\alpha})\sin ka}
{[\sin ka + (k/\tilde{\alpha})\cos ka] - i (k/\tilde{\alpha})\sin ka} ,
\label{eq:Sub-unitario}
\end{equation}
where $\tilde{\alpha}= 2m(u\pm iv)/\hbar^2$ and $k=\sqrt{2mE/\hbar^2}$. It is
easy to see that $\tilde{S}$ is a complex number which can be written in polar
form as in Eq.~(\ref{eq:polar}), where $\theta$ is twice the phase shift plus
$\pi$ and $R=\tilde{S}^{\dagger}\tilde{S}$, is the reflection coefficient; $R<1$
for absorption and $R>1$ for amplification. When $v=0$,
$\tilde{\alpha}=2mu/\hbar^2$ and the unitary case is recovered ($R=1$).

\begin{figure}
\includegraphics[width=\columnwidth]{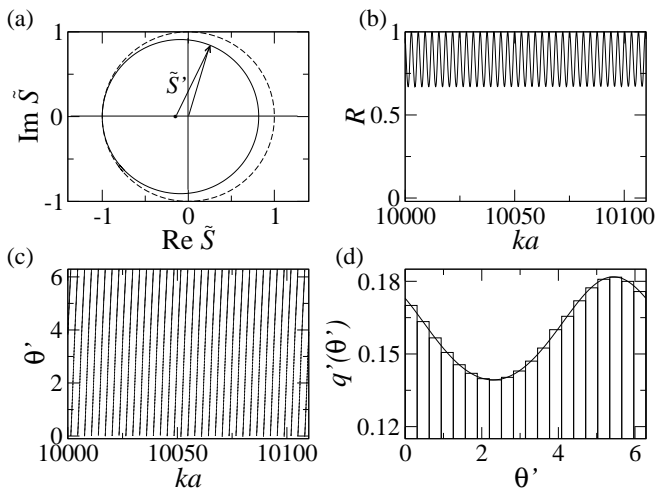}
\caption{(a) In the presence of absorption the motion of $\tilde{S}$ describes
a circle of radius $\sqrt{R_0}<1$ displaced along the real axis with respect to
the unitary case ($v=0$). (b) The reflection coefficient $R$ as a function of
$ka$ takes values between a minimum and unity. (c) The phase $\theta'$ of the
$\tilde{S}'$, seen from the center of the circle, shows the resonances of the
system. (d) The distribution of $\theta'$ shows an excellent agreement with
Poisson's kernel, Eq. (\ref{eq:Poisson-q}).}
\label{fig:grafica1}
\end{figure}

For the potential of Eq.~(\ref{eq:potential}), due to the imaginary part of the
delta potential, the motion of $\tilde{S}$ in the Argand plane describes a
circle of radius $\sqrt{R_0}\neq 1$ displaced along the real axis. In this case
$R$ is not fixed but it is distributed in a certain interval. This circle touch
the unitary one in the point $R=1$ and $\theta=\pi$ where the delta potential is
totally transparent because the wave function has a node just in the position of
the delta potential. The real part of the potential affects only the
distribution of the phase $\theta$ along the circle. To illustrate this, we
present in Fig.~\ref{fig:grafica1} the results for the absorption case with
$ua=va=10^3$ and $ka$ varying from $10^4$ up to obtain 35 resonances (the
amplification case is quite similar). In Fig.~\ref{fig:grafica1}(a) we see that
the motion of $\tilde{S}$ describes a circle of radius $\sqrt{R_0}<1$. In the
panel (b) of the same figure we observe that $R$ varies between a minimum value
$R_{\rm{min}}=(2\sqrt{R_0}-1)^2$ and $R_{\rm{max}}=1$, where  $R_0\approx 0.91$.
If we translate $\tilde{S}$ to $\tilde{S}'=\sqrt{R'}e^{i\theta'}$ such that $R'$
becomes equal to the constant $R_0$, $\theta'$ behaves as shown in
Fig.~\ref{fig:grafica1}(c). What it is interesting here is that the probability
distribution of $\theta'$ is given by Poisson's kernel, as can be seen in
Fig.~\ref{fig:grafica1}(d), where we compare the numerical experiment with the
theoretical result~(\ref{eq:Poisson-q}) with $\overline{\tilde{S}'}$ taken from
the experimental data. Therefore, two relevant parameters are needed, $R_0$ and
$\overline{\tilde{S}'}$. As can be seen, the agreement is excellent.

The probability distribution of $\tilde{S}$ is easily obtained from the one for
$\tilde{S}'$ if we choose $\tilde{S}$ as   
\begin{equation}
\label{eq:traslation}
\tilde{S} = \tilde{S}' - \left( 1 - \sqrt{R_0} \right),
\end{equation}
with $R_0$ the corresponding value for absorption or amplification. From
Eq.~(\ref{eq:Poisson-Rq}) we get 
\begin{eqnarray}
\label{eq:p-cavity}
p(R,\theta) & = & \frac{1}{2\pi} 
\frac{R'(R,\theta)-\left|\overline{\tilde{S}} + 
\left(1-\sqrt{R_0}\right) \right|^2}
{\left|\tilde{S} - \overline{\tilde{S}} \right|^2} 
\nonumber \\ & \times &
\delta\left[R'(R,\theta)-R_0\right],
\end{eqnarray}
where the relevant parameters $R_0$ and $\overline{\tilde{S}}$ are obtained
from the experimental data. To get the marginal distributions, $p(R,\theta)$ of 
Eq.~(\ref{eq:p-cavity}) can be integrated over one of the variables $R$ and
$\theta$. Integration over $\theta$ gives 
\begin{eqnarray}
w(R) & = & \frac{1}{\pi}\, 
\frac{R_0 - \left|\overline{\tilde{S}} + \left(1-\sqrt{R_0}\right) \right|^2}
{\sqrt{\left|(1-R) \left[R-\left(2\sqrt{R_0}-1\right)^2\right]\right|}} 
\nonumber \\ & \times & 
\frac{R+\left(1+\frac{R-\sqrt{R_0}}{1-\sqrt{R_0}}\right)
\textrm{Re}\left(\overline{\tilde{S}}\right) +
\left|\overline{\tilde{S}}\right|^2}
{\left|R+\left(1+\frac{R-\sqrt{R_0}}{1-\sqrt{R_0}}\right)
\overline{\tilde{S}} + {\overline{\tilde{S}}}^{\,2}\right|^2}
\label{eq:w(R)} \\ & \times & 
\left\{
\pm\Theta\left[R-\left(2\sqrt{R_0}-1\right)^2\right] \mp \Theta(R-1)
\right\}, \quad \nonumber
\end{eqnarray}
where $\Theta(x)$ is the Heaviside function, $\textrm{Re}(z)$ stands for the
real part of $z$, and the upper (lower) sign corresponds to absorption
(amplification). 

In similar way, integrating over $R$ the marginal distribution $q(\theta)$ of
$\theta$ can be obtained:
\begin{eqnarray}
q(\theta) & = & \frac{1}{2\pi} \,
\frac{R_0 - \left|\overline{\tilde{S}} + \left(1-\sqrt{R_0}\right) \right|^2}
{\left| \sqrt{R_1(\theta)}\,e^{i\theta} - \overline{\tilde{S}} \right|^2} 
\nonumber \\ &\times &
\frac{\sqrt{R_1(\theta)}}
{\sqrt{R_0 - \left(1-\sqrt{R_0}\right)^2\sin^2\theta}}, 
\label{eq:q(theta)}
\end{eqnarray}
where 
\begin{equation}
\sqrt{R_1(\theta)} = \left(\sqrt{R_0}-1\right)\cos\theta + 
\sqrt{R_0 - \left(1-\sqrt{R_0}\right)^2\sin^2\theta}.
\end{equation}

\begin{figure}
\includegraphics[width=\columnwidth]{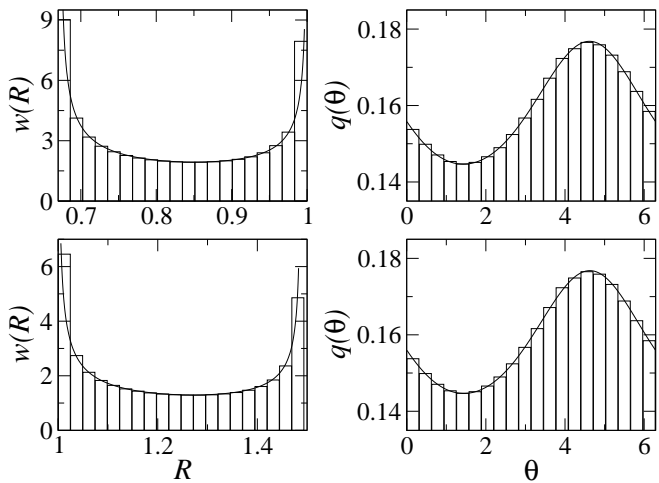}
\caption{The probability distribution of $R$, $w(R)$, is shown on the left
panels for absorption (upper) and amplification (lower). Note that the range of
$R$ is different for absorption than that of amplification, although the
distribution has the same form. The distribution $q(\theta)$ of $\theta$ is
shown on the right panels. It is the same for both cases.}
\label{fig:grafica2}
\end{figure}

The comparison of the marginal distributions $w(R)$ and $q(\theta)$ given in
Eqs.~(\ref{eq:w(R)}) and~(\ref{eq:q(theta)}), with the distributions coming from
the numerical experiment for both, absorption and amplification, are shown in
Fig.~\ref{fig:grafica2} for the same parameters of Fig.~\ref{fig:grafica1}.
Again an excellent agreement is obtained. 

\section{Conclusions}
\label{sec:conclusions}

Extending the analyticity conditions to the non-unitary case, the non-unitary
Poisson kernel was obtained. This was done by adding (substracting) an imaginary
part to the energy for absorbing (amplifying) systems. As a physical
realization, we studied a one-dimensional problem in which the $1\times 1$
scattering matrix is non-unitary. The absorption (amplification) is located just
at one position since the scattering potential was taken as a delta potential
with complex intensity in front of an impenetrable barrier. In this
one-dimensional cavity with local absorption (amplification) the $S$-matrix
moves on a circle with radius smaller (larger) than one; the center of the
circle is not located at the origin of the complex plane, as happens in the
unitary case. In a particular reference framework, such as the one in which the
module of the scattering matrix is constant, the phase of the $S$-matrix is
distributed according to the non-unitary Poisson kernel. The relevant parameters
can be obtained from the numerical simulation and are: (a) the energy average of
the scattering matrix and (b) the reflection coefficient.

We are conscious that the model presented here is somehow artificial and,
therefore, it presents some limitations. For example a delta potential is very
difficult to realize in an experiment. Fortunately, one-dimensional elastic
systems are good candidates to simulate one-dimensional quantum
systems~\cite{Morales}. In this sense the scattering matrix of a rod with a
narrow notch, in which an absorbent foam is added as in Ref.~\cite{Xeridat}, 
could be closely distributed according to the non-unitary Poisson kernel.
Finally, we expect that our one-dimensional model stimulates further
investigations in two-dimensional problems, like chaotic cavities with local
losses or gains. 


\section*{Acknowledgments}

The authors thank to P. A. Mello, G. B\'aez and R. Bernal by useful discussions. This work was supported by CONACyT under project 79613 and by PAPIIT, DGAPA-UNAM under project
IN111311.

\end{document}